# Probing charge transport and background doping in MOCVD grown (010) β-Ga$_2$O$_3$

*Zixuan Feng[1], A F M Anhar Uddin Bhuiyan[1], Zhanbo Xia[1], Wyatt Moore[1], Zhaoying Chen[1], Joe F. McGlone[1], David R. Daughton[2], Aaron R. Arehart[1],*

*Steven A. Ringel[1,3], Siddharth Rajan[1,3] and Hongping Zhao[1,3,\*]*

[1]*Department of Electrical and Computer Engineering, The Ohio State University, Columbus, OH 43210, USA*

[2]*Lake Shore Cryotronics, Westerville, OH 43082, USA*

[3]*Department of Materials Science and Engineering, The Ohio State University, Columbus, OH 43210, USA*

[*] Email: zhao.2592@osu.edu

A new record-high room temperature electron Hall mobility ($\mu_{RT}$ = 194 cm$^2$/V·s at n ~ 8×10$^{15}$ cm$^{-3}$) for β-Ga$_2$O$_3$ is demonstrated in the unintentionally doped thin film grown on (010) semi-insulating substrate via metalorganic chemical vapor deposition (MOCVD). A peak electron mobility of ~9500 cm$^2$/V·s is achieved at 45 K. Further investigation on the transport properties indicate the existence of sheet charges near the epi-layer/substrate interface. Si is identified as the primary contributor to the background carrier in both the epi-layer and the interface, originated from both surface contamination as well as growth environment. Pre-growth hydrofluoric acid cleaning of the substrate lead to an obvious decrease of Si impurity both at interface and in epi-layer. In addition, the effect of MOCVD growth condition, particularly the chamber pressure, on the Si impurity incorporation is studied. A positive correlation between the background charge concentration and the MOCVD growth pressure is confirmed. It is noteworthy that in a β-Ga$_2$O$_3$ film with very low bulk charge concentration, even a reduced sheet charge density can play an important role in the charge transport properties.

Owing to the ultra-wide bandgap energy, Ga$_2$O$_3$ has become the spotlight of semiconductor research for its promising applications in power electronics. Among various polymorphs, β-phase represents the most thermally stable phase with a monoclinic crystal structure and a high breakdown field (~8 MV/cm) projected from its ultra-wide bandgap. [1,2] The significant advantage of β-Ga$_2$O$_3$ as an emerging semiconductor is primarily due to the controllable n-type doping concentration ranging between $10^{16}$ to $10^{20}$ cm$^{-3}$, [3-7] as well as the availability of high-quality native substrate, with 6-inch β-Ga$_2$O$_3$ wafer has been demonstrated recently. [8] The vast effort and fast development of β-Ga$_2$O$_3$ research have enabled various promising device structures that could utilize the material's superb physical properties. For example, β-Ga$_2$O$_3$ lateral metal-semiconductor field-effect transistors (MES-FET) for power radiofrequency (RF) applications have demonstrated 27 GHz cut-off frequency ($F_T$). [9] From the most recent report, vertical fin-structured transistors with normally-off operation have evolved and achieved breakdown voltage (BV) of 1.6 kV. [10] Vertical fin-structured Schottky barrier diode (SBD), has set the record BV of 2.89 kV with Baliga's figure-of-merit (BFOM) of 0.80 GW/cm$^2$ (BV$^2$/R$_{on,sp}$) in DC operation. [11] These advancements in device performance indicate the great potential of β-Ga$_2$O$_3$ in high power electronics, competitive with the traditional wide bandgap semiconductors, such as SiC and GaN.

In the perspective of β-Ga$_2$O$_3$ epitaxy, the key material properties for high BV devices include a) high carrier mobility to minimize the specific on-resistance (R$_{on,sp}$); b) controllable doping at relatively low charge concentration ($10^{14}$ ~ $10^{15}$ cm$^{-3}$); c) thick uniform epi-layers for vertical device structure which could best utilize the high critical field of β-Ga$_2$O$_3$. So far, halide vapor phase epitaxy (HVPE) β-Ga$_2$O$_3$ has demonstrated a fast growth rate (~ 10 μm/hr) on (001)- orientated native substrates with measurable room temperature (RT) electron Hall mobility of 150 cm$^2$/V·s at 2×10$^{15}$ cm$^{-3}$. [3, 12] Low pressure chemical vapor deposition (LPCVD) was demonstrated as a feasible growth method to produce high-quality β-Ga$_2$O$_3$ with controllable n-type doping and tunable growth rates (<1 μm/hr to > 20 μm/hr). [13, 14] For MOCVD β-Ga$_2$O$_3$ epitaxy, most studies have focused on materials grown on (010)-orientated native substrates, with a few reports on (100) orientation. [15-17] So far, MOCVD (010) β-Ga$_2$O$_3$ exhibited high-quality epitaxy with RT Hall mobility > 170 cm$^2$/V·s. [4, 18-20] Our previous work has demonstrated RT mobility value of 184 cm$^2$/V·s on lightly Si-doped (010) thin film with an extracted low-compensation level at N$_A$ < 10$^{15}$ cm$^{-3}$. [4] The full bandgap defect state scanning via deep level transient spectroscopy/deep level optical spectroscopy (DLTS/DLOS) on our MOCVD β-Ga$_2$O$_3$ also confirmed one order lower of total defect density as compared with bulk material or thin films grown by other techniques. [21] Most recently, RT mobility of 130 cm$^2$/V·s and low temperature (LT) peak mobility of 11700 cm$^2$/V·s were reported for MOCVD grown Si-doped β-Ga$_2$O$_3$ thin films. [19] As compared with other wide bandgap semiconductors such as SiC and GaN, the extrodinary high LT mobility in β-Ga$_2$O$_3$ strongly indicates high purity β-Ga$_2$O$_3$ can be achievable via MOCVD. These encouraging results are critical for the development of β-Ga$_2$O$_3$ technology, and MOCVD β-Ga$_2$O$_3$ is fast approaching the theoretically predicted figure-of-merit.

However, typical MOCVD grown β-Ga$_2$O$_3$ exhibited intrinsic n-type background doping. Among the best reported MOCVD β-Ga$_2$O$_3$, the n-type background concentration ranged between 10$^{15}$ cm$^{-3}$ and low-10$^{16}$ cm$^{-3}$. [4, 18-20] The measured background carrier concentration is a result of unintentional n-type doping and low-compensation (N$_A$) level. For high power device applications, controllable doping at low levels (~10$^{14}$ cm$^{-3}$) is desired, which requires to minimize unintentional doping level. In this study, we investigated the background doping in unintentionally-doped (UID) MOCVD β-Ga$_2$O$_3$ via controlled growth condition and wafer preparation.

The UID β-Ga$_2$O$_3$ epitaxy was performed on semi-insulating Fe-doped (010)-orientated native substrates (purchased from Novel Crystal Technology). For the MOCVD growth, Trimethylgallium (TEGa) and O$_2$ were used as precursors, and Ar was used as carrier gas. The group VI/III molar ratio was fixed at 1150 with the growth temperature set at 880 °C. The growth chamber pressure was set at 20, 60, and 100 Torr for the investigation of impurity incorporation. For selected samples, in addition to the regular solvent cleaning (acetone, isopropanol, de-ionized water), the substrates were dipped in HF (5%) for 5 mins, then followed by DI water rinse before loading into the growth chamber. The purpose of this process was to reduce Si contaminants on the substrate surface. Atomic force microscopy (AFM) (Bruker AXS Dimension Icon) was used to characterize the surface morphologies of the as-grown samples. Hall measurements on the epitaxial thin films utilized RT Hall system (Ecopia HMS 3000) and temperature-dependent Hall measurement (Lake Shore CRX-VF probe station equipped with an M91 FastHall controller) with the van der Pauw set up. In order to properly probe the charge transport at the low-temperature range, samples were fabricated with an n$^+$-Ga$_2$O$_3$ MBE regrowth layer prior to the Ti/Au ohmic contact deposition, [22, 23] as shown in **Figure 1(a)**. Ni/Au Schottky contacts were deposited to the vicinity of the ohmic contacts (~100 μm spacing between ohmic and Schottky contacts) to form SBD structures for capacitance-voltage (C-V) measurements. The concentration of Si impurity was quantitatively characterized by secondary ion mass spectroscopy (SIMS).

**Table 1** lists the details of samples grown under different chamber pressure and surface cleaning process. For Sample #1, the substrate was cleaned with regular solvents prior to MOCVD growth, and the growth chamber pressure was set at 60 Torr. The epilayer thickness was 1.2 μm.

RT Hall mobility was measured at 194 cm$^2$/V·s with a carrier concentration of ~8×10$^{15}$ cm$^{-3}$. It showed a slight increase of electron Hall mobility as compared to our previously reported value (184 cm$^2$/V·s at 2.5×10$^{16}$ cm$^{-3}$), [4] which is due to the lower net charge concentration (N$_D$ - N$_A$) and thus less ionized impurity (II) scattering of the carriers. Since the polar-optical phonon (POP) scattering is the dominant factor that limits β-Ga$_2$O$_3$ RT mobility, [24, 25] lowering the charge density would not enhance the Hall mobility significantly, as it has approached the theoretical limit. [24] **Figure 1(b) and 1(c)** plot the temperature-dependent Hall measurement results. The LT peak mobility was measured as 9471 cm$^2$/V·s at 45 K. The experimental data were fitted by modeling taking into account multiple carrier scattering mechanisms, including POP scattering, II scattering, neutral impurity (NI) scattering, and acoustic deformation potential (ADP) scattering. [24, 26] The extracted compensation density was estimated as N$_A$ ~ 7.7×10$^{14}$ cm$^{-3}$. A two-donor model was used to fit the temperature-dependent charge density curve, [26] from which two donor states were identified in the UID film with activation energy E$_{D1}$ ~40 meV and E$_{D2}$ ~150 meV, respectively. The shallow donor state was attributed to Si, which has been widely reported as a shallow donor in β-Ga$_2$O$_3$. The corresponding concentration of N$_{D1}$ was estimated as 8×10$^{15}$ cm$^{-3}$, which also agreed with the SIMS profile, as shown in **Figure 3(a)**. The relatively deeper donor state was estimated with an activation energy of E$_{d2}$ = 150 meV and concentration N$_{d2}$ = 3×10$^{15}$ cm$^{-3}$. Possible origins of the deeper donor state could be from Si on the octahedrally coordinated Ga(II) site, or other types of defects such as antisites or interstitials.

However, there existed a significant peak of Si accumulation at the epi-layer/substrate interface, which added complications to the analysis of charge profile and transport characteristics. For the analysis of temperature-dependent charge transport in Sample #1: 1) By fitting the temperature-dependent Hall mobility with scattering mechanisms [**Figure 1(c)**], the estimated compensation level was at $N_A \sim 7.7 \times 10^{14}$ cm$^{-3}$; however, 2) From the temperature-dependent carrier concentration curve [**Figure 1(b)**], the exponential decay indicated a negligible compensation level. The discrepancy in this analysis can be originated from the interface charges existed at the epi-layer/substrate interface. At temperatures close to room temperature, the charge transport properties were dominated by the charge carriers in the epi-layer, while at low temperatures the interface charges can have a non-negligible contribution to the measured temperature-dependent mobility and carrier concentration. The multicarrier analysis was conducted on Sample #1 from Hall measurement as a function of magnetic field (0.1-2.0 T) at T = 40 K (**Figure 2**). [27] Two electron carriers were extracted from the fitting of the measured conductivity tensor components at different magnetic field, as shown in **Figure 2**. The extraction showed a mobility spectrum of two electron carriers with $N_{s1} = 1.48 \times 10^9$ cm$^{-2}$, $\mu_1 = 10929.7$ cm$^2$/V·s and $N_{s2} = 1.73 \times 10^9$ cm$^{-2}$, $\mu_2 = 8274.6$ cm$^2$/V·s ($N_s$ represents the integrated sheet charge concentration of each carrier component). The first carrier component can represent the transport of the top epi-layer with low impurity, and the second component can correspond to the charges at the growth interface. Note that there may exist other impurities at the interface that can affect the sheet charge transport. Fe impurity represents one of the compensators that was originated from the Fe-doped $Ga_2O_3$ substrates or surface contamination. [4, 28] Therefore, the analysis of charge transport properties in low-doped β-$Ga_2O_3$ thin films requires specific consideration of the contribution from impurities/defects at the interface.

The influence from the interface charges was also confirmed from the electrical characterization for samples with thinner epitaxial layers. RT Hall measurements were performed for UID β-$Ga_2O_3$ samples with epi-layer of 200 nm and 400 nm grown on Fe-doped (010) substrates. RT Hall mobility of $\mu = 112$ cm$^2$/V·s and 122 cm$^2$/V·s were measured for these two films with sheet charge density of $N_s \sim 2\times10^{12}$ cm$^{-2}$ and $3\times10^{12}$ cm$^{-2}$, respectively. The reduced RT mobility was due to the non-negligible impact from the interface charge transport.

In order to reduce the Si impurity concentration at the epi-layer/substrate interface, HF cleaning process was used to remove the SiO$_x$ component on the substrate surface. Sample #2, #3, #4 as listed in **Table 1** were all dipped in diluted HF (5%) for 5 mins prior to the MOCVD growth. As shown by the SIMS profiles in **Figure 3(a) and 3(b)**, for Sample #1 and #3 grown under identical conditions, we observed three times (3×) reduction of total Si impurity level near the interface by integrating the SIMS concentration profiles. Moreover, from RT Hall measurements, the total charge concentration in the HF-cleaned sample was reduced to half as compared to that of the sample without cleaning. The HF cleaning procedure reduced the Si impurity concentration near the growth interface as well as the Si riding into the epi-layer. [17, 20] However, the measured RT mobility of Sample #3 was 173 cm$^2$/V·s. The slight reduction in mobility value as compared to Sample #1 can be due to comprimised material quality from the surface HF cleaning process.

MOCVD growth pressure was identified as a critical parameter that affects background Si incorporation. Growth experiments were conducted with chamber pressure at 20, 60, and 100 Torr, respectively. C-V charge profiles in **Figure 4(b)** exhibited a positive correlation between the growth pressure and net charge density ($N_D - N_A$) in the epi-layer. As the growth pressure decreased, Si impurity incorporation reduced. Considering the Schottky metal-semiconductor junction depletion, the depletion depths were hundreds of nm at zero bias due to the low net charge concentration. The extracted charge profiles from CV measurements showed a similar trend as the SIMS Si profile, where the charge density increased going towards the growth interface.

For Sample #2 grown at 20 Torr, the C-V characteristic (**Figure 4(a)**) exhibited very low unit capacitance and the obvious existence of sheet charge at the interface. This agreed well with the SIMS profile shown in **Figure 3(c)** that this sample had low Si in epi-layer (below SIMS detection limit) as well as the reduced Si concentration peak at the interface. Note that the measured RT mobility of Sample #2 was at 128 cm$^2$/V·s, which was similar to the values measured from the samples with thin UID epi-layers. This clearly indicates the electrical measurement was influenced by the interface charges in addition to the bulk charges in the epi-layer. In order to accurately characterize the charge transport properties in epi-layer with extremely low background doping, the influence from Si contamination at the interface needs to be eliminated by approaches such as intentional doping with compensators e.g., Fe or Mg. Alternatively, similar to the method used in GaAs, in which low-temperature GaAs buffer layer with semi-insulating property was used in device applications, [29-31] intrinsically insulating Ga$_2$O$_3$ buffer may be grown with certain conditions.

Surface morphologies of the as-grown samples were characterized by AFM to evaluate the effects from HF cleaning process as well as MOCVD growth pressure, as shown in **Figure 5**. All samples were grown with targeted similar film thickness (~ 1.2 μm). Comparing Sample #1 and #3, the surface root-mean-square (RMS) roughness increased from 2.2 nm to 4.17 nm with HF cleaning, which indicated the cleaning process may introduce epi-surface modification. As growth pressure increased, the surface roughness showed a monotonically increase from 3.41 nm (P=20 Torr), 4.17 nm (P=60 Torr), to 5.67 nm (P=100 Torr). The trend indicates that surface roughness increases as the growth pressure increases, which can be due to the reduced surface desorption of Ga adatoms and limited surface diffusion at higher O$_2$ partial pressure. [32-34] For Sample #1 without HF cleaning, it exhibited the most uniform morphology with fine elongated grain structures aligned along the [001] orientation.

In summary, charge transport properties in MOCVD grown (010) β-Ga$_2$O$_3$ films were investigated. Close to theoretical limit RT Hall mobility of 194 cm$^2$/V·s was achieved in UID film with intrinsic carrier concentration n ~ 8×10$^{15}$ cm$^{-3}$. The corresponding peak LT mobility was measured as 9471 cm$^2$/V·s at 45K. Si impurity was identified as the primary shallow donor that contributed to the background conductivity in UID films. In addition, Si contamination at the growth interface can play an important role in the charge transport analysis especially for films with low carrier concentrations. Neglecting this factor can lead to the misinterpretation of experimental data obtained from electrical characterization. HF cleaning of the Ga$_2$O$_3$ substrates can effectively reduce Si concentration at the interface. However, it also led to an increase of surface roughness and reduced electron mobility. MOCVD growth pressure was identified as a key parameter that affects the background Si incorporation. Si impurity incorporation decreased as the growth pressure reduced. <u>To accurately probe the epi-films with low background doping potentially achievable by low growth pressure, the elimination of the substrate interface charges is required.</u> Results from this work show promises to achieve controllable n-type doping in MOCVD grown β-Ga$_2$O$_3$ at low 10$^{15}$ cm$^{-3}$.

**Acknowledgements**

The authors acknowledge the funding support from the Air Force Office of Scientific Research FA9550-18-1-0479 (AFOSR, Dr. Ali Sayir), and the National Science Foundation (1810041).

**References:**

1. M. Higashiwaki, K. Sasaki, A. Kuramata, T. Masui, and S. Yamakoshi, *Appl. Phys. Lett.* **2012**, *100*, 013504.
2. R. Roy, V.G. Hill and E.F. Osborn, *J. Am. Chem. Soc.* **1952**, *74*, 719.
3. K. Goto, K. Konishi, H. Murakami, Y. Kumagai, B. Monemar, M. Higashiwaki, A. Kuramata, and S. Yamakoshi, *Thin Solid Films* **2018**, *666*, 182.
4. Z. Feng, A F M A. U. Bhuiyan, M. R. Karim, and H. Zhao, *Appl. Phys. Lett.* **2019**, *114*, 250601.
5. K. Sasaki, A. Kuramata, T. Masui, E. G. Víllora, K. Shimamura, and S.Yamakoshi, *Appl. Phys. Express* **2012**, *5*, 035502.
6. S. Rafique, L. Han, A. T. Neal, S. Mou, J. Boeckl, and H. Zhao, *Phys. Status Solidi A* **2018**, *215*, 1700467.
7. K. D. Leedy, K. D. Chabak, V. Vasilyev, D. C. Look, J. J. Boeckl, J. L. Brown, S. E. Tetlak, A. J. Green, N. A. Moser, A. Crespo, D. B. Thomson, R. C. Fitch, J. P. McCandless, and G. H. Jessen, *Appl. Phys. Lett.* **2017**, *111*, 012103.
8. A. Kuramata, presented at IWGO Conf. Recent Progress in Edge-Defined Film-Fed Growth and Halide Vapor Phase Epitaxy of β- $Ga_2O_3$ for Power Device Applications, Columbus, Ohio, August, **2019**.
9. Z. Xia, H. Xue, C. Joishi, J. McGlone, N. K. Kalarickal, S. H. Sohel, M. Brenner, A. Arehart, S. Ringel, S. Lodha, W. Lu, and S. Rajan, *IEEE Electr. Device L.* **2019**, *40*, 1052.
10. Z. Hu, K. Nomoto, W. Li, R. Jinno, T. Nakamura, D. Jena, and H. Xing, in *Proceedings of the 31st International Symposium on Power Semiconductor Devices & ICs*, 1.6 kV Vertical $Ga_2O_3$ FinFETs With Source-Connected Field Plates and Normally-off Operation, Shanghai, China, May, **2019**.
11. W. Li, K. Nomoto, Z. Hu, D. Jena, and H. G. Xing, IEEE Electr. Device L. 41, 107 (2020).
12. K. Nomura, K. Goto, R. Togashi, H. Murakami, Y. Kumagai, A. Kuramata, S. Yamakoshi, and A. Koukitu, *J. Cryst. Growth* **2014**, *405*, 19.
13. S. Rafique, M. R. Karim, J. M. Johnson, J. Hwang, and H. Zhao, *Appl. Phys. Lett.* **2018**, *112*, 052104.
14. Z. Feng, M. R. Karim, and H. Zhao, *APL Mater.* **2019**, *7*, 022514.
15. G. Wagner, M. Baldini, D. Gogova, M. Schmidbauer, R. Schewski, M. Albrecht, Z. Galazka, D. Klimm, and R. Fornari, *Phys. Status Solidi A* **2014**, *211*, 27.
16. A. Fiedler, R. Schewski, M. Baldini, Z. Galazka, G. Wagner, M. Albrecht, and K. Irmscher, *J. Appl. Phys.* **2017**, *122*, 165701.
17. R. Schewski, K. Lion, A. Fiedler, C. Wouters, A. Popp, S. V. Levchenko, T. Schulz, M. Schmidbauer, S. Bin Anooz, R. Grüneberg, Z. Galazka, G. Wagner, K. Irmscher, M. Scheffler, C. Draxl, and M. Albrecht, *APL Mater.* **2019**, *7*, 022515.
18. Y. Zhang, F. Alema, A. Mauze, O. S. Koksaldi, R. Miller, A. Osinsky, and J. S. Speck, *APL Mater.* **2019**, *7*, 022506.

19. F. Alema, Y. Zhang, A. Osinsky, N. Valente, A. Mauze, T. Itoh, and J. S. Speck, *APL Mater.* **2019**, *7*, 121110.
20. F. Alema, Y. Zhang, A. Osinsky, N. Orishchin, N. Valente, A. Mauze, and J. S. Speck, *APL Mater.* **2020**, *8*, 021110.
21. H. Ghadi, J. F. McGlone, C. M. Jackson, E. Farzana, Z. Feng, A F M A. U. Bhuiyan, H. Zhao, A. R. Arehart, and S. A. Ringel, *APL Mater.* **2020**, *8*, 021111.
22. Z. Xia, C. Joishi, S. Krishnamoorthy, S. Bajaj, Y. Zhang, M. Brenner, S. Lodha, and S. Rajan, *IEEE Electr. Device L.* **2018**, *39*, 568.
23. Y. Zhang, C. Joishi, Z. Xia, M. Brenner, S. Lodha, and S. Rajan, *Appl. Phys. Lett.* **2018**, *112*, 233503.
24. N. Ma, N. Tanen, A. Verma, Z. Guo, T. Luo, H. Xing, and D. Jena, *Appl. Phys. Lett.* **2016**, *109*, 212101.
25. Y. Kang, K. Krishnaswamy, H. Peelaers, and C. G. Van de Walle, J. *Phys.: Condens. Matter* **2017**, *29*, 234001.
26. A. T. Neal, S. Mou, S. Rafique, H. Zhao, E. Ahmadi, J. S. Speck, K. T. Stevens, J. D. Blevins, D. B. Thomson, N. Moser, K. D. Chabak, and G. H. Jessen, *Appl. Phys. Lett.* **2018**, *113*, 062101.
27. J. S. Kim, D. G. Seiler, and W. F. Tseng, *J. Appl. Phys.* **1993**, *73*, 8324.
28. A. Mauze, Y. Zhang, T. Mates, F. Wu, and J. S. Speck, *Appl. Phys. Lett.* **2019**, *115*, 052102.
29. A. C. Warren, J. M. Woodall, J. L. Freeouf, D. Grischkowsky, D. T. McInturff, M. R. Melloch, and N. Otsuka, *Appl. Phys. Lett.* **1990**, *57*, 1331.
30. D. C. Look, D. C. Walters, M. O. Manasreh, J. R. Sizelove, C. E. Stutz, and K. R. Evans, *Phys. Rev. B* **1990**, *42*, 3578.
31. X. Liu, A. Prasad, W. M. Chen, A. Kurpiewski, A. Stoschek, Z. Liliental‐Weber, and E. R. Weber, *Appl. Phys. Lett.* **1994**, *65*, 3002.
32. M. Baldini, M. Albrecht, A. Fiedler, K. Irmscher, R. Schewski, and G. Wagner, *ECS J. Solid State Sci. Technol.* **2017**, *6*, Q3040.
33. K. Sasaki, M. Higashiwaki, A. Kuramata, T. Masui, and S. Yamakoshi, *J. Cryst. Growth* **2014**, *392*, 30.
34. P. Mazzolini, P. Vogt, R. Schewski, C. Wouters, M. Albrecht, and O. Bierwagen, *APL Mater.* **2019**, *7*, 022511.

**Table Caption**

**Table 1.** Summary of the UID samples with different growth chamber pressure and surface preparation. Samples were all grown with targeted thickness around ~1.2 μm with the same precursor flow rate. The VI/III ratio was constant at 1150 and the growth temperature was 880 °C.

* The epi-layer is mostly semi-insulating as inferred from C-V charge depth profile. Mobility measured from Hall is mostly from the interface charge near the epi-layer/substrate interface.

**Figure Caption**

**Figure 1.** (a) Schematics for the n+ MBE regrowth contacts, (b) Measured and calculated temperature-dependent carrier density for Sample #1, and (c) carrier mobility as a function of temperature for Sample #1. Dot represents the measured data. Dashed lines denote the fitted value based on calculation.

**Figure 2.** Experimental data for the conductivity tensor components $\sigma_{xx}$ and $\sigma_{xy}$ (red dots) and QMSA-fitted curve (blue curve) as a function of applied magnetic field for Sample #1 at T = 40 K.

**Figure 3.** SIMS depth profiles of Si impurity in UID homoepitaxy samples. Si peaks at depth around 1.1 μm are the epi-layer/substrate interface, Sample #1 (a) grown at 60 Torr without HF cleaning process, Sample #3 (b) grown at 60 Torr with HF cleaning before growth, Sample #2 (c) grown at 20 Torr with HF cleaning before growth.

**Figure 4.** C-V characteristics (a) of all four samples measured from the SBD at excitation frequency of 100 kHz. Extracted charge profiles (b) ($N_d$-$N_a$) versus depletion depth of all four samples.

**Figure 5.** Surface AFM images on the as-grown surface of all samples grown at different chamber pressure and substrate surface preparation.

Table 1.

| Sample | Surface Preparation | Growth Chamber Pressure (Torr) | RT Bulk Carrier Concentration (cm$^{-3}$) | RT Hall Mobility (cm$^2$/V·s) |
|---|---|---|---|---|
| #1 | Solvent | 60 | 7.7×10$^{15}$ | 194 |
| #2 | Solvent, HF | 20 | Semi-insulating* | 128* |
| #3 | Solvent, HF | 60 | ~3.9×10$^{15}$ | 173 |
| #4 | Solvent, HF | 100 | ~1.2×10$^{16}$ | 160 |

Figure 1.

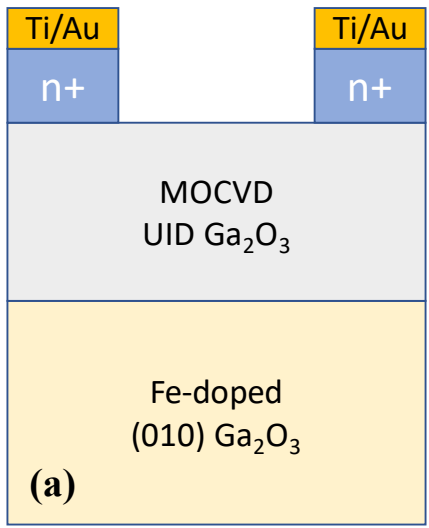
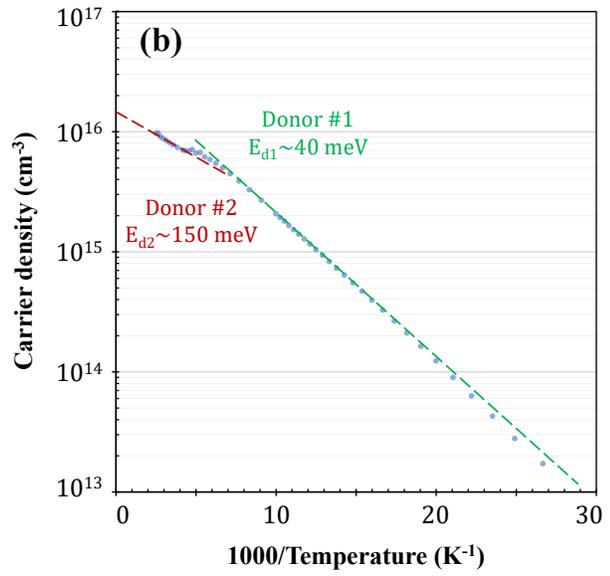
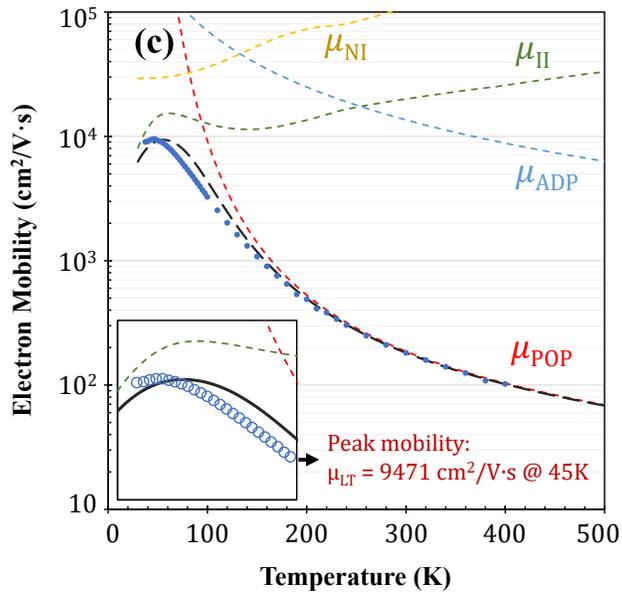

Figure 2.

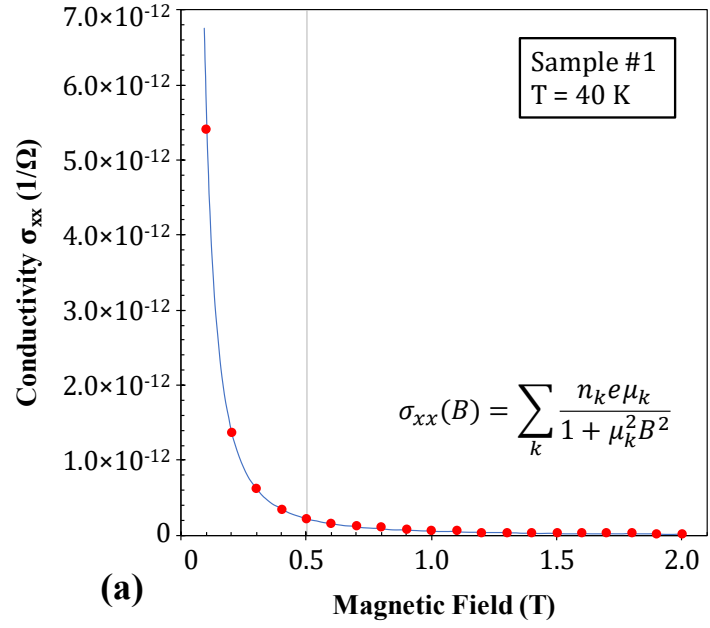

(a)

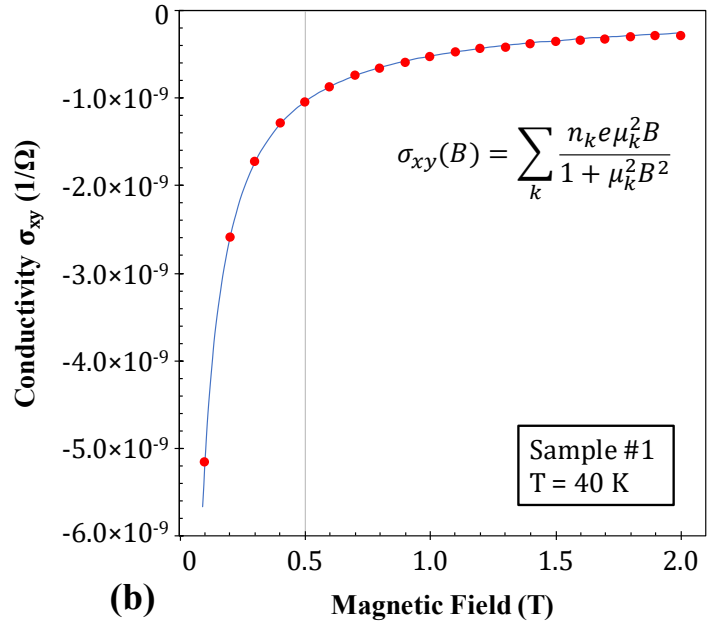

(b)

Figure 3.

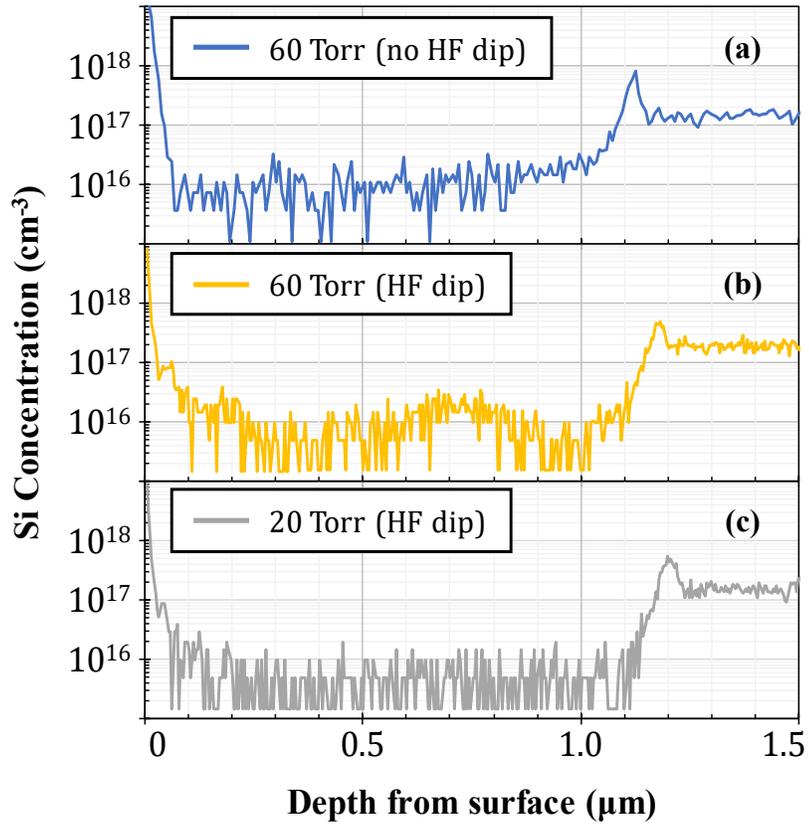

Figure 4.

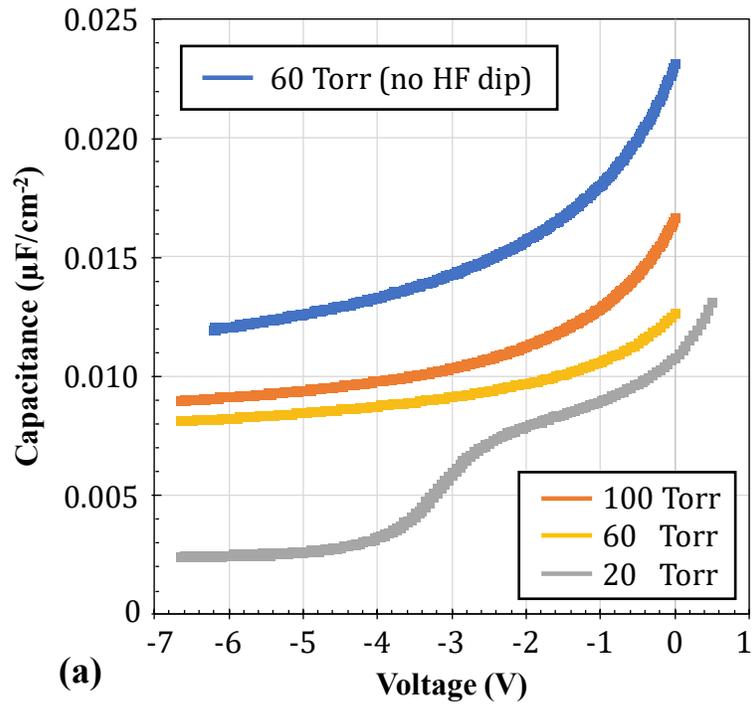

(a)

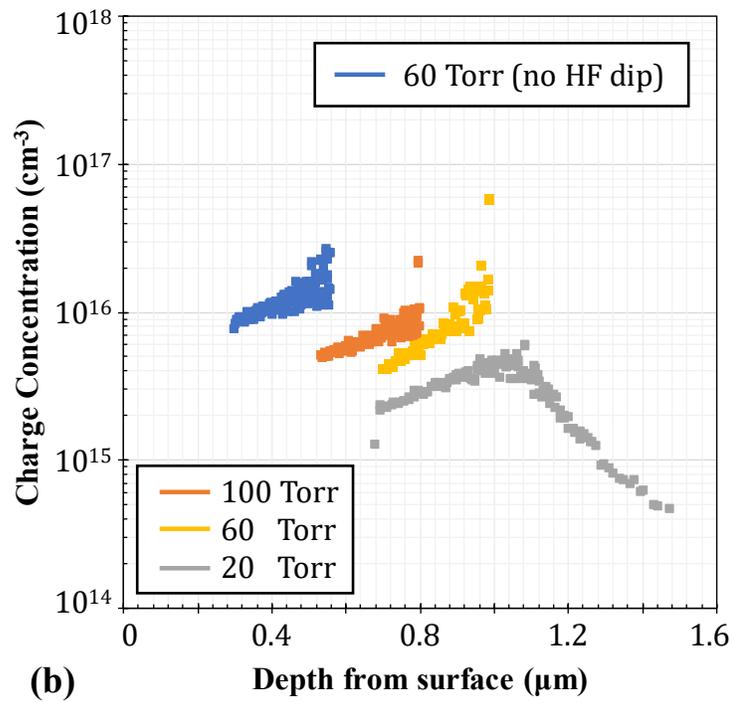

(b)

Figure 5.

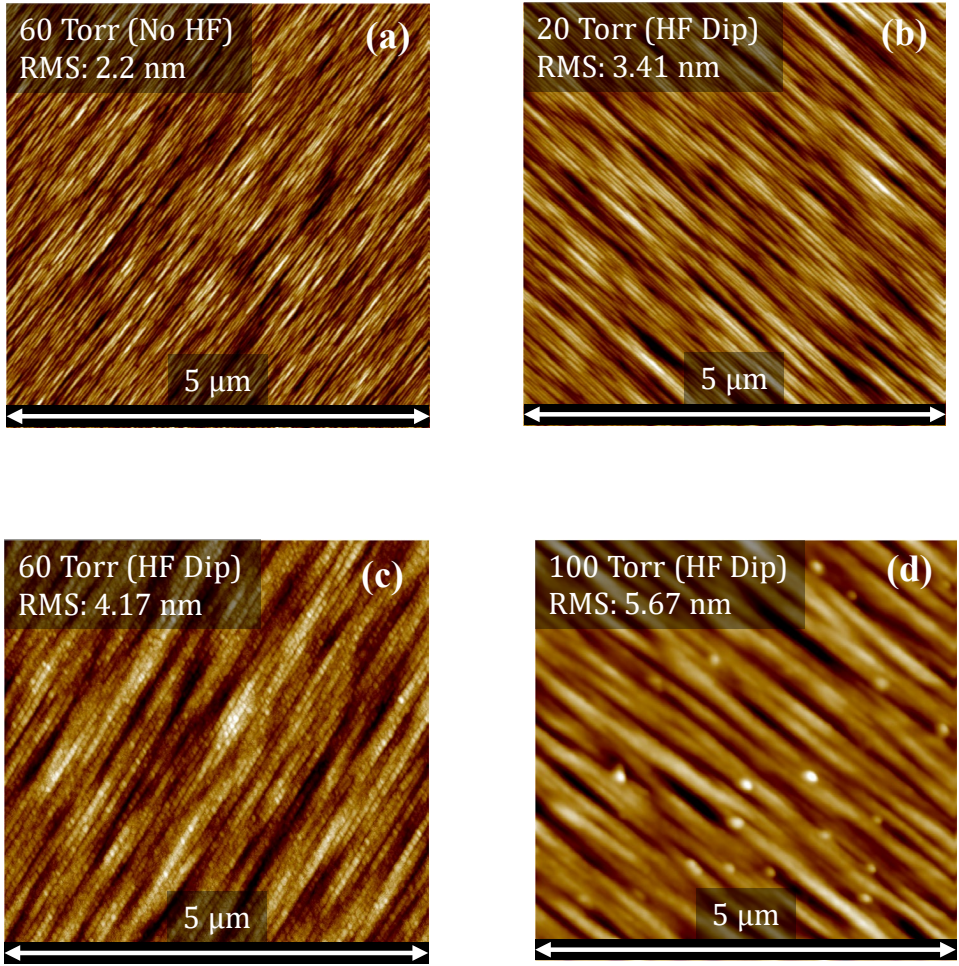